\documentclass[aps,prb,twocolumn,showpacs,superscriptaddress]{revtex4}

\usepackage{graphicx} 
\usepackage{soul}
\usepackage{multirow}
\usepackage[T1]{fontenc}
\usepackage[latin9]{inputenc}
\usepackage{rotating}
\usepackage{amsmath}
\usepackage{amssymb}
\usepackage{tabularx}

\begin{document}

\title{Structures of Fluorinated Graphenes and Their Signatures}

\author{H. \c{S}ahin}
\affiliation{UNAM-Institute of Materials Science and
Nanotechnology, Bilkent University, 06800 Ankara, Turkey}

\author{M. Topsakal}
\affiliation{UNAM-Institute of Materials Science and
Nanotechnology, Bilkent University, 06800 Ankara, Turkey}

\author{S. Ciraci}\email{ciraci@fen.bilkent.edu.tr}
\affiliation{UNAM-Institute of Materials Science and
Nanotechnology, Bilkent University, 06800 Ankara, Turkey}
\affiliation{Department of Physics, Bilkent University, 06800 Ankara,
Turkey}

\date{\today}

\begin{abstract}

Recent synthesis of fluorinated graphene introduced 
interesting stable derivatives of graphene. In particular, 
fluorographene (CF), namely fully fluorinated chair 
conformation, is found to display crucial features, such as 
high mechanical strength, charged surfaces, local magnetic 
moments due to vacancy defects and a wide band gap rapidly 
reducing with uniform strain. These properties, as well as 
structural parameters and electronic densities of states are 
found to scale with fluorine coverage. However, most of the 
experimental data reported to date neither for CF, nor for 
other C$_n$F structures complies with the results obtained 
from first-principles calculations. In this study, we attempt 
to clarify the sources of disagreements.

\end{abstract}


\maketitle

\section{Introduction}

Active research on graphene\cite{novo1} revealed not only 
numerous exceptional properties 
\cite{geim,berger,kats,hasan2010} but also have prepared a 
ground for the discovery of several graphene based materials. 
Preparation of freestanding graphene sheets with nonuniform 
oxygen coverage have been achieved.\cite{dikin} More
recently the synthesis of two-dimensional hydrocarbon in 
honeycomb structure, so called \textit{graphane} 
\cite{novo-graphane} (CH), showing diverse electronic, magnetic 
and mechanical properties
\cite{sofo,boukhvalov,graphane-h1,graphane-h2,graphane-m} is 
reported.

According to Pauling scale F has electronegativity of 3.98, which is 
higher than those of C(2.55), H(2.20) and O(3.44), and hence 
fluorination of graphene is expected to result in a material, 
which may be even more interesting than both graphene oxide 
and CH. Before the first synthesis of graphene, fluorinated 
graphite has been treated theoretically.
\cite{charlier,takagi} Owing to promising properties
revealed for CH, fluorinated graphene structures are now
attracting considerable interest
\cite{boukhvalov2,cheng,robinson,nair,yeni1, yeni2, yeni3, 
yeni4} despite uncertainties in their chemical compositions 
and atomic structures. In an effort to identify the 
structures of fluorinated samples, previous theoretical 
models attempted to deduce the lowest energy structures.
\cite{charlier,boukhvalov2} In addition, band gaps of 
different structures calculated within Density Functional 
Theory (DFT) are compared with the values revealed
through specific measurements.\cite{robinson,nair} However,
neither the stability of proposed structures are questioned, 
nor underestimation of band gaps within DFT has been a 
subject matter. Raman spectrum by itself, has been limited in 
specifying C$_n$F structures.\cite{nair}

In this work, we first determined stable C$_n$F structures 
for $n \leq 4$. Then we revealed specific properties (such as 
internal structural parameters, elastic constants, formation 
and binding energies, energy band gap and photoelectric 
threshold) for those stable structures as signatures to 
identify the derivatives probed experimentally. We placed an 
emphasis on fully fluorinated graphene or fluorographene 
(CF), in which D and G Raman peaks of bare graphene disappear 
after long fluorination period.\cite{robinson,nair} Present 
study reveals that the properties, such as structural 
parameters, binding energy, band gap and phonon modes of 
various fluorinated structures are strongly dependent on the 
binding structure of F atoms and their composition. Some of 
these properties are found to roughly scale with F coverage. 
While the stable C$_2$F chair structure is metallic, CF is a 
nonmagnetic insulator with a band gap, $E_{g}$, being much 
larger than 3 eV, i.e. a value attributed experimentally to 
fully fluorinated graphene. In view of the calculated 
diffusion constant, Raman active modes and other properties, 
available experimental data suggest that domains (or grains) 
of various C$_n$F structures with extended and imperfect 
grain boundaries can coexist after fluorination process. 
Hence the measured properties are averaged from diverse 
perfect and imperfect regions.

\section{Computational Methodology}

Our predictions are obtained from first-principles plane wave
calculations\cite{vasp} within DFT, which is demonstrated to 
yield rather accurate results for carbon based materials. 
Calculations are performed using spin-polarized local density 
approximation (LDA)\cite{lda} and projector augmented wave 
(PAW) potentials.\cite{paw} Kinetic energy cutoff $ \hbar^2 |
\mathbf{k}+\mathbf{G}|^2 / 2m $ for plane-wave basis set is 
taken as 500 eV. In the self-consistent potential and total 
energy calculations of fluorographene a set of (25x25x1) 
\textbf{k}-point sampling is used for Brillouin zone (BZ) 
integration. The convergence criterion of self consistent 
calculations for ionic relaxations is $10^{-5}$ eV between 
two consecutive steps. By using the conjugate gradient 
method, all atomic positions and unit cells are optimized
until the atomic forces are less than 0.03 eV/\AA. Pressures 
on the lattice unit cell are decreased to values less than 
0.5 kBar. The energy band gap, which is usually 
underestimated in DFT, is corrected by frequency-dependent 
GW$_{0}$ calculations.\cite{gw} In GW$_{0}$  corrections 
screened Coulomb potential, W, is kept fixed to initial DFT 
value W$_{0}$ and Green's function, G, is iterated four 
times. Various tests are performed regarding vacuum spacing, 
kinetic energy cut-off energy, number of bands, \textbf{k}-points 
and grid points. Finally, the band gap of CF is found 
7.49 eV after GW$_{0}$ correction, which is carried out by 
using (12x12x1) \textbf{k}-points in BZ, $15$~\AA~ vacuum 
spacing, default cut-off potential for GW$_{0}$, 192 bands 
and 64 grid points. Phonon frequencies and phonon 
eigenvectors are calculated using the Density Functional
Perturbation Theory (DFPT).\cite{pwscf}

\section{Structures of fluorinated graphene}
Each carbon atom of graphene can bind only one F atom and 
through coverage (or decoration) of one or two sides of 
graphene, one can achieve diverse C$_{n}$F structures. 
Uniform F coverage is specified by $\Theta = 1/n$ (namely one 
F adatom per $n$ C atoms), whereby $\Theta = 0.5$ corresponds 
to half fluorination and $\Theta = 1$ is fluorographene CF. 
The adsorption of a single F atom to graphene is precursor 
for fluorination. When placed at diverse sites of a $(4 
\times 4)$ supercell of graphene, simple F atom moves to the 
top site of a carbon atom and remains adsorbed there. For the 
resulting structure spin-polarized state (with 0.4 Bohr magneton) is only 2 meV favorable than the nonmagnetic state. The binding energy of F is 
$E_b$=2.71 eV in equilibrium, that is a rather strong binding 
unlike many other adatoms adsorbed to graphene. An energy 
barrier, $Q_{B} =\sim$0.45 eV, occurs along its minimum 
energy migration path. Our calculations related with the 
minimum energy path of a single F atom follow hexagons of 
underlying graphene. Namely, F atom migrates from the highest 
binding energy site, i.e. top site (on top of carbon atom) 
to the next top site through bridge site (bridge position 
between two adjacent carbon atoms of graphene). The 
corresponding diffusion constant for a single F atom, $D= \nu 
a e^{-Q_{B}/k_{B}T}$ is calculated in terms of the lattice 
constant, $a=2.55$ \AA~and characteristic jump frequency 
$\nu\approx$39 THz. Experiments present evidences
that energy barriers on the order of 0.5 eV would make the
adatoms mobile.\cite{nair,ref1} Moreover, this energy barrier is 
further lowered even it is collapsed in the presence of a 
second F atom at the close proximity. Consequently, this 
situation together with the tendency towards clustering 
favors that C$_n$F grains (or domains) of different $n$  
on graphene can form in the course of fluorination. We note 
that the energy barrier for the diffusion of a single carbon 
adatom adsorbed on the bridge sites of graphene was 
calculated to be in the similar energy range. Carbon adatoms on 
graphene were found to be rather mobile. That energy barrier 
for single C adatom was found to decrease, even to collapse at 
the close proximity of a second adatom.\cite{carbon-adatom}

\begin{figure}
\centering
\includegraphics[width=8.5cm]{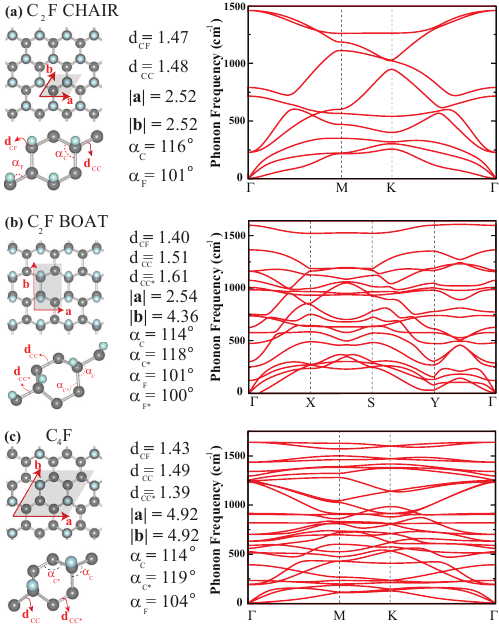}
\caption{(Color online) Atomic structure and calculated 
phonon bands (i.e. phonon frequencies versus wave vector, \textbf{k}) of various optimized C$_n$F
structures calculated along the symmetry directions of BZ. 
Carbon and fluorine atoms are indicated by black (dark) and 
blue (light) balls, respectively. (a) C$_2$F Chair structure. 
(b) C$_2$F Boat structure. (c) C$_4$F structure. Units are 
\AA~for structural parameters and cm$^{-1}$ for frequencies.}

\label{fig1}
\end{figure}

\begin{table*}
\caption{Comparison of the calculated properties of four 
stable, fluorinated graphene structures (namely CF, C$_2$F 
chair, C$_2$F boat and C$_4$F) with those of graphene and CH. 
Lattice constant, $a=b$ ($a \neq b$ for rectangular lattice); 
C-C bond distance, $d_{CC}$ (second entries with slash differ 
from the previous one); C-X bond distance (X indicating H (F) 
atom for CH (CF)), $d_{CX}$; the buckling, $\delta$; angle 
between adjacent C-C bonds, $\alpha_{C}$; angle between 
adjacent C-X and C-C bonds, $\alpha_{X}$; total energy per 
cell comprising 8 carbon atoms $E_T$; formation energy per X 
atom relative to graphene, $E_f$; binding energy per X atom
relative to graphene, $E_b$ (the value in parenthesis 
$E_{b^{'}}$ excludes the X-X coupling); desorption energy, 
$E_d$ (see the text for formal definitions); energy band gap 
calculated by LDA, $E_{g}^{LDA}$; energy band gap corrected 
by GW$_{0}$, $E_{g}^{GW_{0}}$; photoelectric threshold, 
$\Phi$; in-plane stiffness, $C$; Poison ratio, $\nu$. All 
materials are treated in hexagonal lattice, except C$_2$F 
boat having rectangular lattice.}

\label{table} \centering{}\begin{tabular}{lcccccccccccccccc}
\hline \hline

Material  & $a$ $(b)$  & $d_{CC}$  & $d_{CX}$  & $\delta$  &
$\alpha_{C}$ & $\alpha_{X}$  & $E_{g}^{LDA}$  & $E_{g}^{GW_{0}}$ &
$E_{T}$  & $E_{f}$  &$E_{b}$$(E_{b^{'}})$& $E_{d}$& $\Phi$& $C$ &
$\nu$ \tabularnewline

  & (\AA{})  &(\AA{})  & (\AA{})  & (\AA{})  &
(deg)  & (deg)  & ($eV$)  &($eV$)  &  ( $eV$ )  & ($eV$)  & ($eV$)
& ($eV$) &($eV$)&($J/m^{2}$)  & \tabularnewline \hline

Graphene\cite{h-ansiklopedi}  & 2.46  &1.42  & -  & 0.00  & 120
& -  & 0.00  & 0.00  &-80.73  & - & - & - & 4.77  & 335  & 0.16
\tabularnewline \hline

CH\cite{graphane-h1} &2.51  & 1.52  & 1.12  & 0.45  & 112  & 107 &
3.42 & 5.97 & -110.56 & 0.39  & 2.8(2.5) & 4.8 &4.97  & 243  & 0.07
\tabularnewline \hline

CF   & 2.55  & 1.55  & 1.37  & 0.49  & 111  & 108 & 2.96 &
7.49  & -113.32 & 2.04  & 3.6(2.9) & 5.3 &7.94  & 250  & 0.14
\tabularnewline \hline

C$_2$F chair & 2.52  & 1.48  & 1.47  & 0.29  & 116  & 101  & metal
& metal  & -89.22  & 0.09& 1.7(0.9) & 1.2 &8.6/5.6  & 280  & 0.18
\tabularnewline \hline

C$_2$F boat & 2.54(4.36) & 1.51/1.61 & 1.40 & 0.42 & 114/118 &
100/101 & 1.57 & 5.68 & -92.48 & 0.91 & 2.5(1.6) & 2.4 &7.9/5.1 &
286(268)  & 0.05   \tabularnewline \hline

C$_{4}$F & 4.92  & 1.49/1.39  & 1.43  & 0.34  & 114/119 & 104
& 2.93 & 5.99  & -87.68  &1.44 & 3.0(2.7)& 3.5 & 8.1/5.6 & 298 & 0.12
\tabularnewline \hline \hline
\end{tabular}
\end{table*}

\begin{figure}
\centering
\includegraphics[width=8.5cm]{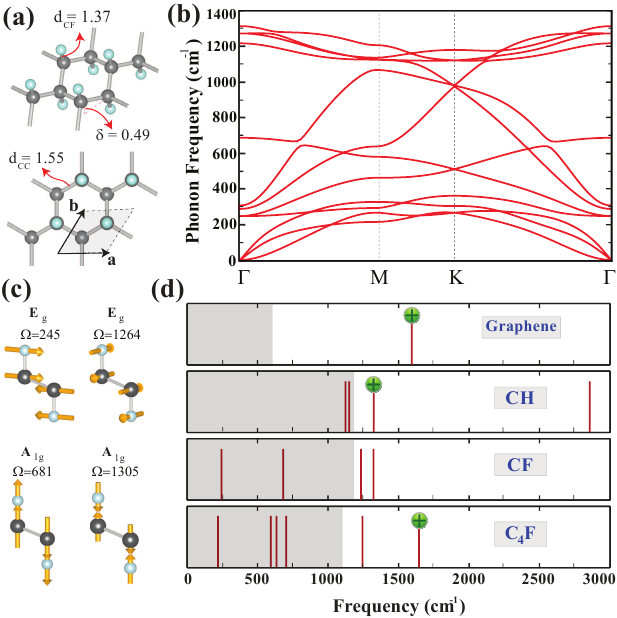}
\caption{(Color online) (a) Atomic structure of  
fluorographene CF. $a$ and $b$ are the lattice vectors ($|a|
=|b|$) of hexagonal structure; $d_{CC}$ ($d_{CF}$) is the C-C 
(C-F) bond distance; $\delta$ is the buckling. (b) Phonon 
frequencies versus wave vector \textbf{k} of optimized CF calculated along 
symmetry directions in BZ. (c) Symmetries, frequencies and 
descriptions of Raman active modes of CF. (d) Calculated 
Raman active modes of graphene, CH, CF and C$_4$F are 
indicated on the frequency axis. Those modes indicated
by "+" are observed experimentally. There is no experimental
Raman data in the shaded regions. Units are \AA~for 
structural parameters and cm$^{-1}$ for frequencies.}
\label{fig2}
\end{figure}

In earlier theoretical studies,
\cite{charlier,boukhvalov2,robinson}
the total energies and/or binding energies were taken as 
criteria for whether a given C$_n$F structure exists. Even if 
a C$_n$F structure seems to be in a minimum on the Born-
Oppenheimer surface, its stability is meticulously examined 
by calculating frequencies of all phonon modes in BZ. Here we 
calculated phonon dispersions of most of optimized C$_n$F 
structures. We found C$_4$F, C$_2$F boat, C$_2$F chair (See
Fig.~\ref{fig1}) and CF chair (See Fig.~\ref{fig2}) 
structures have positive frequencies throughout the Brillouin 
zone indicating their stability.

Some of phonon branches of C$_n$F structures (for example, CF 
boat) have imaginary frequencies and hence are unstable, in 
spite of the fact that their structures can be optimized. The 
possibility that these unstable structures can occur at 
finite and small sizes is, however, not excluded. For stable 
structures, the gap between optical and acoustical branches 
is collapsed, since the optical branches associated with the 
modes of C-F bonds occur at lower frequencies. This situation 
is in contrast with the phonon spectrum of graphane,
\cite{graphane-h1} where optical modes related with C-H bonds 
appear above the acoustical branches at $\sim$ 2900 
cm$^{-1}$.

The formation energy of fluorination is defined as 
$E_f=(n_{F_{2}} E_{T,F_{2}}+E_{T,Gr}-E_{T,C_{n}F})/n_F$ in 
terms of the total ground state energies of optimized 
structures of graphene and fluorinated graphenes at different 
composition, respectively, $E_{T,Gr}$, $E_{T,C_{n}F}$, and 
the total ground state energy of single carbon atom, 
$E_{T,C}$, of F$_2$ molecule and F atom, $E_{T,F_{2}}$ and 
$E_{T,F}$. Similarly, the binding energy of F atom relative 
to graphene including F-F coupling is $E_{b}=(E_{T,Gr}+n_{F} 
E_{T,F}- E_{T,C_{n}F})/n_{F}$ and without F-F coupling 
$E_{b^{'}}=(E_{T,Gr}+E_{T,n_{F}F}- E_{T,C_{n}F})/n_{F}$. Here 
$E_{T,n_{F}F}$ is the total energy of suspended single or 
double layers of F occupying the same positions as in C$_n$F. 
The desorption energy, $E_d$ is the energy required to remove 
one single F atom from the surface of C$_n$F. $n_{F_{2}}$ and 
$n_F$ are numbers of F$_2$ molecules and F atoms, 
respectively. The total energies are calculated in 
periodically repeating supercells comprising 8 carbon atoms 
and keeping all the parameters of calculations described 
above using spin-polarized as well as spin unpolarized LDA. 
Lowest (magnetic or nonmagnetic) total energy is used as 
ground state total energy.

Fluorographene (CF), where F atoms are bound to each C atom 
of graphene alternatingly from top and bottom sides is 
energetically most favorable structure. Upon full 
fluorination, the planar honeycomb structure of C atoms 
becomes buckled (puckered) and C-C bond length increases by 
$\sim$10\%. At the end, while planar $sp^2$-bonding of 
graphene is dehybridized, the buckled configuration is 
maintained by $sp^3$-like rehybridization. In 
Table~\ref{table}, the calculated lattice constants, internal
structural parameters, relevant binding energies and energy 
band gaps of stable C$_n$F structures are compared with those 
of bare graphene and CH.\cite{graphane-h1} Notably, internal 
parameters (such as $\delta$, C-C bond length) as well as 
lattice constants of various C$_n$F structures vary with F 
coverage, $\Theta$. CF, has highest values for $E_f$, $E_b$, 
$E_{b^{'}}$ and $E_d$ given in Table~\ref{table}; those of 
C$_4$F are second highest among stable C$_n$F structures.

Since Raman spectrum can convey information for a particular
structure and hence can set its signature, calculated Raman 
active modes of stable C$_4$F and CF structures together with 
those of graphene and CH are also indicated in 
Fig.~\ref{fig2} (c) and (d). It is known that the only 
characteristic Raman active mode of graphene at 1594 
cm$^{-1}$ is observed so far.
\cite{r-graphene} Similarly, for CH the mode at $\sim$1342 
cm$^{-1}$ is observed.\cite{novo-graphane} One of two
Raman active modes of C$_4$F at 1645 cm$^{-1}$ seems to be 
observed.\cite{robinson} In compliance with the theory,
\cite{ferrari} phonon branches of all these observed modes 
exhibit a kink structure. However, none of the Raman active
modes of CF revealed in Fig.~\ref{fig2} has been observed 
yet. Raman spectroscopy in the low frequency range may be 
useful in identifying experimental structures.

\section{Electronic Structures}

\begin{figure}
\centering
\includegraphics[width=8.5cm]{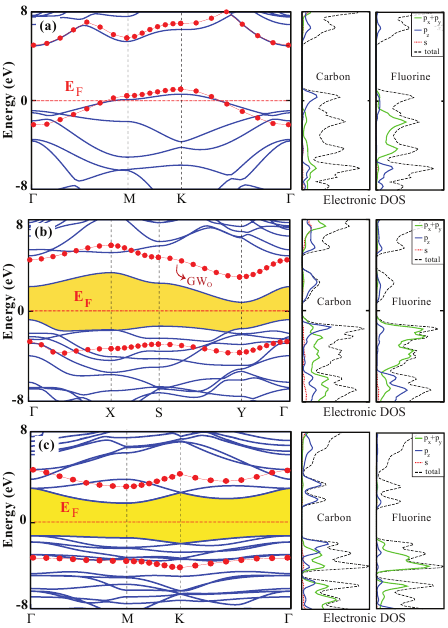}
\caption{(Color online) Energy band structures of various 
stable C$_n$F structures together with the orbital projected 
densities of states and the total densities of states (DOS).
The LDA band gaps are shaded and the zero of energy is set to 
the Fermi level $E_{F}$. Total DOS is scaled to 45\%. Valence 
and conduction band edges after GW$_{0}$ correction are 
indicated by filled/red circles. (a) C$_2$F chair structure. 
(b) C$_2$F Boat structure. (c) C$_4$F structure.}
\label{fig3}
\end{figure}

\begin{figure}
\centering
\includegraphics[width=8.5cm]{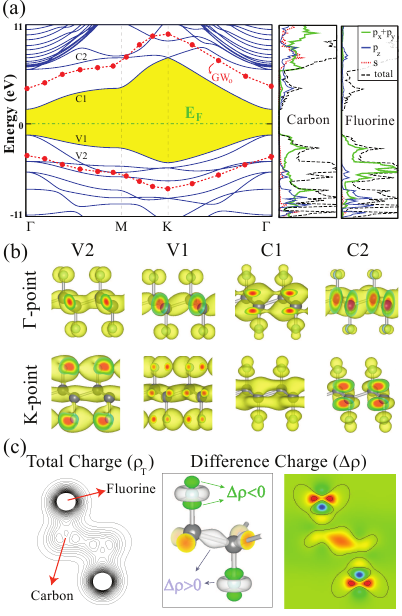}
\caption{(Color online) (a) Energy band structure of CF 
together with the orbital projected DOS and 
total densities of states. The LDA band gap is shaded and the 
zero of energy is set to the Fermi level, $E_{F}$. Valence
and conduction band edges after GW$_{0}$ correction are 
indicated by filled/red circles. (b) Isosurfaces of charge 
densities of states corresponds to first (V1), second (V2) 
valance and first (C1) and second (C2) conduction bands at 
the $\Gamma$- and $K$-points. (c) Contour plots of the total 
charge density $\rho_T$ and difference charge density 
$\Delta\rho$ in the plane passing through F-C-C-F atoms. 
Contour spacings are 0.03 e/\AA$^{3}$.}

\label{fig4}
\end{figure}

Energy bands, which are calculated for the optimized
C$_4$F, C$_2$F boat, C$_2$F chair and CF chair structures are
presented in Fig.~\ref{fig3} and Fig.~\ref{fig4} structures, 
respectively. The orbital projected densities of states (PDOS) 
together with the total densities of states of these 
optimized structures are also presented. Analysis of the 
electronic structure can also provide data to reveal the 
observed structure of fluorinated graphene. As seen in
Table~\ref{table}, stable C$_n$F structures have LDA band 
gaps ranging from 0 eV to 2.96 eV. Surprisingly, C$_2$F chair 
structure is found to be a metal owing to the odd number of 
valance electrons in the primitive unit cell. Even if various 
measurements on the band gap of fluorinated graphene lie in 
the energy range from 68 meV\cite{cheng} to 3 eV,\cite{nair} 
these calculated band gaps are underestimated by LDA. 
Incidentally, the band gaps change significantly after they 
are corrected by various self-energy methods. In fact, the 
correction using GW$_0$ self-energy method predicts a rather 
wide band gap of 7.49 eV for CF. The corrected band gaps for 
C$_2$F boat structure and C$_4$F are 5.68 eV and 5.99 eV, 
respectively. It should be noted that the GW$_0$ self-energy 
method has been successful in predicting the band gaps of 3D
semiconductors.\cite{hse06}

While predicting much larger band gap for CF, the measured 
band gap of $\sim$3 eV reported by Nair et al.\cite{nair} 
marks the serious discrepancy between theory and experiment.
The character of the band structure of CF are revealed from 
the analysis of projected density of states as well as charge
densities of specific bands in Fig.~\ref{fig4} (b). 
Conduction band edge consists of the antibonding combination 
of $p_{z}$-orbitals of F and C atoms. The $p_{z}$-orbitals of 
C atoms by themselves, are combined to form $\pi$-bands. The 
bands at the edge of the valence band are derived from the 
combination of C-$p_{x}+p_{y}$ and F-$p_{x}+p_{y}$ orbitals. 
The total contribution of C orbitals to the valence band can 
be viewed as the contribution of four tetrahedrally 
coordinated $sp^{3}$-like hybrid orbitals of $s$- and
$p$-orbitals of C atoms. However, the deviation from
tetrahedral coordination increases when $n$ increases or 
single side is fluorinated. As a matter of fact, the total 
density of states presented in Fig.~\ref{fig3} and 
Fig.~\ref{fig4} mark crucial differences. In this respect, 
spectroscopy data is expected to yield significant 
information regarding the observed structures of fluorinated
graphenes.

The contour plots of the total charge density, $\rho_T$, in 
the F-C-C-F plane suggests the formation of strong covalent 
C-C bonds from the bonding combination of two C-$sp^{3}$ 
hybrid orbitals. The difference charge density, $\Delta \rho$ 
(which is obtained by subtracting the charges of free C and 
free F atoms situated at their respective positions in
 CF), indicates charge transfer to the middle of C-C bond and 
to F atom, revealing the bond charge between C atoms and 
ionic character of C-F bond. However, the value of charge 
transfer is not unique, but diversifies among different 
methods of analysis.\cite{charge-analysis} Nevertheless, the 
direction of calculated charge transfer is in compliance with 
the Pauling ionicity scale and is corroborated by calculated 
Born effective charges, which have in-plane ($\parallel$) and 
out-of-plane ($\perp$) components on C atoms, $Z^{*}_{C,
\parallel}=0.30$, $Z^{*}_{C,\perp}=0.35$ and on F atoms
$Z^{*}_{F,\parallel}=-0.30$, $Z^{*}_{F,\perp}=-0.35$.

Finally, we note that perfect CF is a nonmagnetic insulator. 
However, a single isolated F vacancy attains a net magnetic 
moment of 1 Bohr magneton ($\mu_{B}$) and localized defect 
states in the band gap. Creation of an unpaired $\pi$-
electron upon F vacancy is the source of magnetic moment. 
However, the exchange interaction between two F-vacancies 
calculated in a (7x7x1) supercell is found to be nonmagnetic 
for the first nearest neighbor distances due to spin pairings. 
Similar to graphane,\cite{graphane-h1,graphane-h2} it is also 
possible to attain large magnetic moments on F-vacant domains 
in CF structures.

\begin{figure*}
\centering
\includegraphics[width=14cm]{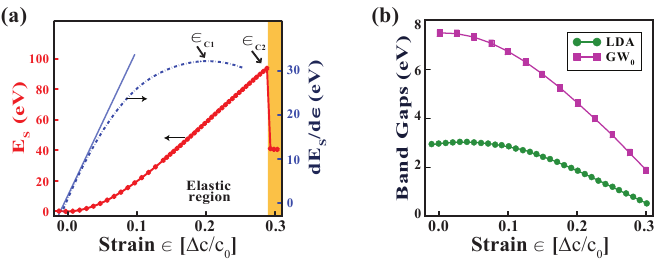}
\caption{(Color online) (a) Variation of strain energy and 
its first derivative with respect to the uniform strain 
$\epsilon$. Orange/gray shaded region indicates the plastic 
range. Two critical strains in the elastic range are labeled 
as $\epsilon_{c1}$ and $\epsilon_{c2}$. (b) Variation of the 
band gaps with $\epsilon$. LDA and $GW_0$ calculations are 
carried out using 5x5 supercell having the lattice parameter 
of c$_{0}$=5a, and $\Delta c$ is its stretching}

\label{figs5}
\end{figure*}

\section{Elastic Properties of CF}

Having analyzed the stability of various C$_n$F structures
with $n=$1,2 and 4, we next investigate their mechanical
properties. The elastic properties of this structure can
conveniently be characterized by its Young's modulus and 
Poison's ratio. However, the in-plane stiffness $C$ is known 
to be a better measure of the strength of single layer 
honeycomb structures, since the thickness of the layer $h$ 
cannot be defined unambiguously. Defining $A_{0}$ as the 
equilibrium area of a C$_n$F structures, the in-plane 
stiffness is obtained as $C=(\delta E^{2}_{s}/ \delta 
\epsilon^2)/A_{0}$, in terms of strain energy $E_s$
and uniaxial strain $\epsilon$.\cite{graphane-m} The values
of in-plane stiffness $C$, and Poisson's ratio $\nu$, 
calculated for stable C$_n$F structures are given in 
Table~\ref{table} together with the values calculated for 
graphene and graphane. For example, the calculated values of 
CF are $C$=250 J/m$^2$ and $\nu$=0.14. It is noted that $C$ 
increases with $n$. For CF (i.e. $n=1$), the in-plane 
stiffness is close to that calculated for CH. It appears that 
either the interaction between C-F bonds in CF (or the 
interaction between C-H bonds in CH) does not have 
significant contribution to the in-plane stiffness. The main
effect occurs through dehybridization of $sp^2$ bonds of 
graphene through the formation C-F bonds (or C-H bonds).

A value of Young's modulus around 0.77 $TPa$ can be 
calculated by estimating the thickness of CF as $h$=3.84 
\AA{}, namely the sum of the thickness of graphene (3.35
\AA) and buckling, $\delta$ (0.49 \AA). This value is
smaller, but comparable with the value proposed for graphene, 
i.e. $\sim$ 1 $TPa$. Here the contribution of C-F bonds to
the thickness of CF is neglected, since the interaction 
between C-F bonds has only negligible effects on the strength 
of CF.

In Fig.~\ref{figs5} the variation of strain energy $E_s$ and 
its derivative, $\delta E_{s}/\delta \epsilon$, with strain, 
$\epsilon$ are presented in both elastic and plastic
regions. Two critical strain values, $\epsilon_{c_{1}}$ and
$\epsilon_{c_{2}}$, are deduced. The first one,
$\epsilon_{c_{1}}$, is the point where the derivative curve
attains its maximum value. This means that the structure can 
be expanded under smaller tension for higher values of 
strain. This point also corresponds to phonon 
instability\cite{graphane-m} where the longitudinal acoustic 
modes start to become imaginary for $\epsilon > 
\epsilon_{c_{1}}$. The second critical point, 
$\epsilon_{c_{2}}$  $(\simeq 0.29$), corresponds to the 
yielding point. Until this point the honeycomb like structure 
is preserved, but beyond it the plastic deformation sets in. 
We note that for $\epsilon_{c_{1}}<\epsilon<\epsilon_{c_{2}}$ 
the system is actually in a metastable state, where the 
plastic deformation is delayed. Under the long wavelength 
perturbations, vacancy defects and high ambient temperature 
$\epsilon_{c_{2}}$ approaches to $\epsilon_{c_{1}}$. In fact, 
our further molecular dynamics simulations show that 
$\epsilon_{c_{2}} \rightarrow$ 0.17 at 300K and to 0.16 at 
600K. In the presence of periodically repeating F-vacancy and 
C+F-divacancy, the value of $\epsilon_{c_{2}}$ is also 
lowered to 0.21 and 0.14, respectively. Apart from phonon
instability occurring at high $\epsilon$, the band gap is 
strongly affected under uniform expansion. In 
Fig.~\ref{figs5}(b) we show the variation of LDA and $GW_0$-
corrected band gaps under uniform expansion. The LDA gap 
slightly increases until $\epsilon=0.05$ and then decreases 
steadily with increasing $\epsilon$. The $GW_0$-corrected 
band gap essentially decreases with increasing strain. For 
example, its value decreases  by 38\% for
$\epsilon=0.20$.

\section{Conclusions}

Present analysis of fluorinated graphenes shows that 
different C$_n$F structures can form at different level of
F coverage. Calculated properties of these structures, such 
as lattice parameter, $d_{CC}$ distance, band gap, density 
of states, work function, in plane stiffness $C$, Poisson's
ratio and surface charge, are shown to depend on $n$ or 
coverage $\Theta$. Relevant data reported in various 
experiments do not appear to agree with the properties 
calculated any one of the stable C$_n$F structures. This 
finding lets us to conclude that domains of various C$_n$F 
structures can form in the course of the fluorination of 
graphene. Therefore, the experimental data may reflect a 
weighted average of diverse C$_n$F structures together with 
extended defects in grain boundaries. In this respect, 
imaging of fluorinated graphene surfaces by scanning 
tunneling and atomic force microscopy, as well as x-ray 
photoemission spectroscopy is expected to shed light on the 
puzzling inconsistency between theory and experiment.

Finally, our results show a wide range of interesting 
features of C$_n$F structures. For example, a perfect CF 
structure as described in Fig.~\ref{fig2} is a stiff, 
nonmagnetic wide band gap nanomaterial having substantial 
surface charge, but attains significant local magnetic moment 
through F-vacancy defects. Moreover, unlike graphane, half 
fluorinated graphene with only one side
fluorinated is found to be stable, which can be further
functionalized by the adsorption of adatoms to other side. 
For example, hydrogen atoms adsorbed to other side attain 
positive charge and hence permanent transversal electric 
field, which can be utilized to engineer electronic 
properties.

\section{ACKNOWLEDGMENTS}

This work is supported by TUBITAK through Grant No:108T234. 
Part of the computational resources has been provided by 
UYBHM at ITU through grant Grant No. 2-024-2007. We thank the 
DEISA Consortium (www.deisa.eu), funded through the EU FP7 
project RI-222919, for support within the DEISA Extreme 
Computing Initiative. S. C. acknowledges the partial support 
of TUBA, Academy of Science of Turkey. The authors would also 
like to acknowledge the valuable suggestions made by D. Alfe.

\end{document}